\documentclass{bioinfo}
\usepackage{subfigure}
\usepackage{stfloats}
\usepackage{amsmath}
\usepackage{multirow}
\usepackage[ruled,vlined]{algorithm2e}
\copyrightyear{2012}
\pubyear{2012}

\begin{document}

\firstpage{1}

\title[Optimal Interval Intersection Counting]{
Binary Interval Search (BITS):  \\
A Scalable Algorithm for Counting Interval Intersections}

\author[Layer \textit{et~al}]
{Ryan M. Layer$^1$, 
Kevin Skadron\,$^1$,
Gabriel Robins$^1$, 
Ira M. Hall$^2$,
and
Aaron R. Quinlan$^3$\footnote{to whom correspondence should be addressed}}

\address{$^{1}$Department of Computer Science, University of Virginia,
Charlottesville, VA\\
$^{2}$Department of Biochemistry and Molecular Genetics, University of Virginia, Charlottesville, VA\\
$^{3}$Department of Public Health Sciences and Center for Public Health
Genomics, University of Virginia, Charlottesville, VA}

\history{Received on XXXXX; revised on XXXXX; accepted on XXXXX}

\editor{Associate Editor: XXXXXXX}

\maketitle

\begin{abstract}
\section{Motivation:}
The comparison of diverse genomic datasets is fundamental to
understanding genome biology.
Researchers must explore many large datasets of genome intervals (e.g., genes,
sequence alignments) to place their experimental results in a
broader context and to make new discoveries.  Relationships between genomic
datasets are typically measured by identifying intervals
that intersect: that is, they overlap and thus share a common genome interval.
Given the continued advances in DNA sequencing technologies, efficient methods
for measuring statistically significant relationships between many sets of 
genomic features is crucial for future discovery.

\section{Results:}
We introduce the Binary Interval Search (BITS) algorithm, a novel and
scalable approach to interval set intersection. We
demonstrate that BITS outperforms existing methods at counting interval 
intersections. Moreover, we show that BITS is intrinsically suited to parallel 
computing architectures such as Graphics Processing Units (GPUs) by illustrating 
its utility for efficient Monte-Carlo simulations measuring the significance of 
relationships between sets of genomic intervals.

\section{Availability:}
\href{https://github.com/arq5x/bits}{https://github.com/arq5x/bits}

\section{Contact:} arq5x@virginia.edu
\end{abstract}

\section{Introduction}
Searching for intersecting intervals in multiple sets of genomic features is
crucial to nearly all genomic analyses. For example, interval intersection is
used to compare ChIP enrichment between experiments and cell types, identify
potential regulatory targets, and compare genetic variation among many
individuals.  Interval intersection is the fundamental operation in a broader
class of ``genome arithmetic'' techniques, and as such, it underlies
the functionality found in many genome analysis software packages~\citep{kent2002,giardine2005,li2009,quinlan2010}.

As high throughput sequencing technologies have become the 
\emph{de facto} molecular tool for genome biology, there is an acute need for 
efficient approaches to interval intersection. Microarray
techniques for measuring gene expression and chromatin
states have been largely supplanted by sequencing-based techniques, 
and whole-exome and whole-genome
experiments are now routine. Consequently, most genomics labs now conduct
analyses including datasets with billions of genome intervals.
Experiments of this size require substantial computation time per pair-wise
comparison; moreover, typical analyses require comparisons to many large sets of
genomic features. Existing approaches scale poorly and
are already reaching their performance limits. We therefore argue
the need for new scalable algorithms to allow discovery to keep pace with the
scale and complexity of modern datasets.

In this manuscript, we introduce the Binary Interval Search (BITS) algorithm as
a novel and scalable solution to the fundamental problem of counting the number
of intersections between two sets of genomic intervals.  BITS uses two binary searches (one each for start and end coordinates) to identify intersecting intervals. As such, our algorithm executes
in $\Theta(N \log N)$ time, where $N$ is the number of intervals, which can be
shown to be optimal for the interval intersection counting problem by a
straight-forward reduction to element uniqueness (known to be
$\Theta(N\log N)$~\citep{misra1982}).  We illustrate that a sequential version 
of BITS outperforms existing approaches. We also demonstrate that BITS 
is intrinsically suited to parallel architectures. The parallel 
version performs the same amount of work as the sequential version 
(i.e., there is no overhead) which means the algorithm is work-efficient, and 
because each parallel thread performs equivalent work, BITS has little thread 
divergence. While thread divergence degrades performance on any architecture 
(finished threads must wait for over-burdened threads to complete),
the impact is particularity acute for GPUs where threads share a program counter
and any divergent instruction must be executed on every thread.

% , in the  sequential
% case, BITS outperformed both BEDTools and UCSC in all tests, and performed
% particularly well for larger datasets and for common (e.g., exome sequencing,
% RNAseq, and ChIPseq) genomic applications.  Most importantly, we show that  the
% BITS algorithm is well suited for parallel execution.  The parallel version
% performs the same amount of work as the sequential version (i.e., there is no
% overhead) which means the algorithm is work-efficient, and because each parallel
% thread performs roughly the same amount of work, the algorithm has little thread
% divergence. As such the BITS algorithm is a particularly promising approach for
% applications requiring large numbers of intersection operations, such as massive
% database searches and Monte Carlo-based  simulations to measure the
% significance of relationships between sets of genome features.

        %%%%%%%%%%%%%%%%%%%%%%%%%%%%%%%%%%%%%%%%%%%%%
        % INTRO: The Interval Set Intersection problem
        %%%%%%%%%%%%%%%%%%%%%%%%%%%%%%%%%%%%%%%%%%%%%
\subsection{The Interval Set Intersection problem}
We begin by reviewing some basic definitions.  A genomic \emph{interval} is a 
continuous stretch of a genome with a chromosomal start and end location 
(e.g., a gene), and a genomic \emph{interval set} is a collection of genomic 
intervals (e.g., all known genes).  
% More generally, an interval $a$ is the set of all
% numbers between a start value and an end value, which can be represented as 
% the pair $(a.start, a.end)$.  
Two intervals $a$ and $b$ {\em intersect} when 
$(a.start \leq b.end)$ and $(a.end \geq b.start)$.  The intersection of two
interval sets $A=\{a_1, a_2, \dots, a_N\}$ and $B=\{b_1, b_2, \dots, b_M\}$ is
the set of interval pairs:
\vspace{-.75em}
\begin{equation*}
	\begin{split}
		\mathcal{I}(A,B)= \{ \langle a,b \rangle |& a \in A, b \in B, \\
		& a.start \leq b.end \wedge a.end \geq b.start\}
	\end{split}
\end{equation*}
Intervals within a set can intersect, but {\em self-intersections} are not
included in $\mathcal{I}(A,B)$.  
% The interval set intersection problem is a
% special case of the segment intersection problem~\citep{bentley1979} where all
% points are located on the same line, and each segment belongs to one of two
% sets. 
There are four natural sub-problems for interval set intersection: 1) 
\emph{decision} - does there exist at least one 
interval in $A$ that intersects an interval in $B$?; 2) {\em counting} - how 
many total intersections exist between sets $A$ and $B$?; 3) {\em 
per-interval counting} - how many intervals in $B$ intersect each interval in 
$A$?; 4) {\em enumeration} - what is the set of each pair-wise interval 
intersections between $A$ an $B$? While BITS solves all 
four sub-problems, it is designed to efficiently \emph{count} the number of 
intersections between two sets, and as such, it excels at the \emph{decision}, 
\emph{counting}, and\emph{per-interval counting} problems.

% \begin{enumerate}
%   \item The {\em decision problem}
%   $\mathcal{I_D}(A,B)$:  given interval sets $A$ and $B$, does there exist at
%   least one interval in $A$ that intersects an interval in $B$?
% 
%   \item The {\em counting problem}
%   $\mathcal{I_C}(A,B)$: how many pair-wise intersections exist between the
%   intervals $A$ and $B$? 
% 
%   \item The {\em per-interval counting problem}
%   $\mathcal{I_P}(A,B)$: how many intervals in $B$ intersect each
%   interval in $A$?
% 
%   \item The {\em enumeration problem}
%   $\mathcal{I}(A,B)$: what is the set of pair-wise interval intersections
%   between $A$ an $B$?
% \end{enumerate}

%%%%%%%%%%%%%%%%%%%%%%%%%%%%%%%%%%%%%%%%%%%%%%%%%%%%%%%%%%%
% INTRO: Parallelization limitations of existing approaches
%%%%%%%%%%%%%%%%%%%%%%%%%%%%%%%%%%%%%%%%%%%%%%%%%%%%%%%%%%%
\subsection{Limits to parallelization}

Interval intersection has many applications in genomics, and as such, several
algorithms have been developed that, in general, are either based on
trees~\citep{kent2002,alekseyenko2007}, or linear sweeps of
pre-sorted intervals~\citep{richardson2006}.

The UCSC genome browser introduced a widely-used scheme based on
R-trees. This approach partitions intervals from one dataset into
hierarchical ``bins''.  Intervals from a second dataset are then compared 
to matching bins (not the entire dataset) to narrow the search
for intersections to a focused portion of the genome.  While this 
approach is used by the UCSC Genome Browser, BEDTools~\citep{quinlan2010},
and SAMTOOLS~\citep{li2009}, the algorithm is inefficient for counting 
intersections since all intervals in each candidate bin must be
\emph{enumerated} in order to count the intersections.  Since the number of
intersections is at most quadratic, any enumeration-based algorithm is
$O(N^2)$.
 
Moreover, these existing approaches are poor candidates for parallelization.
Thread divergence can be a significant problem for hierarchical binning methods.  
If intervals are not uniformly distributed
(e.g., exome sequencing or RNA-seq), then a small number of bins will contain
many intervals while most other bins are empty. 
Consequently, threads searching full
bins will take substantially longer than threads searching empty bins.  In
contrast, BITS counts intersections directly without enumerating
intersecting intervals and therefore, the underlying interval distribution does
not impact the relative workload of each thread.

% Binning the hierarchical binning data structure can also be difficult to build
% in some parallel architectures.  For example, in NVIDIA's CUDA at least one
% extra passes over the interval database is required to allocate the correct
% amount of space for each bin.  BITS uses integer arrays, which easily map to
% most architectures.

Recent versions of BEDTools and BEDOPS~\citep{neph2012} conduct a
linear ``sweep'' through pre-sorted datasets while maintaining an auxiliary
data structure to track intersections as they are encountered. While the
complexity of such sequential sweep algorithms is theoretically optimal, the 
amount of parallelism that exists is limited, and some overhead is required to
guarantee correctness.  Any linear sweep algorithm must maintain the ``sweep
invariant''~\citep{mckenney2009}, which states that all segment starts, ends, 
and intersections behind the sweep must be known.  A parallel sweep algorithm
must either partition the input space such that each section can be swept in
parallel without violating the invariant, or threads must communicate 
about intervals that span partitions.  In the first case parallelism is limited
to the number of partitions that can be created, and threads can diverge when 
the number of intervals in each partition are unbalanced.  In the second case,
the communication overhead between threads prevents work efficiency and can 
have significant performance implications.  In BITS, the amount of parallelism
depends only on the number of intervals and not the distribution of intervals
within the input space, and there is no communication between threads.

% Several parallel sweep algorithms have been proposed~\citep{goodrich1993,
% kriegel1991, mckenney2009} for segment intersection (a generalization of
% interval intersection), but they all depend on a partitioned input space
% scheme and thus have limited parallelism and increased overhead.

%%%%%%%%%%%%%%%%%%%%%%%%%%%%%%%%%%%%%%%%%%%%%%%%%%%%%%%%
% METHODS
%%%%%%%%%%%%%%%%%%%%%%%%%%%%%%%%%%%%%%%%%%%%%%%%%%%%%%%%
\section{Methods}

%%%%%%%%%%%%%%%%%%%%%%%%%%%%%%%%%%%%%%%%%%%%%%%%%%%%%%%%
% METHODS: BITS
%%%%%%%%%%%%%%%%%%%%%%%%%%%%%%%%%%%%%%%%%%%%%%%%%%%%%%%%

A seemingly facile method for finding the intersection of $A$ and $B$ would be
to treat one set, $A$, as a ``query'' set, and the other, $B$, as a
``database''.  If all of the intervals in the database were sorted by their
starting coordinates, it would seem that binary searches could be used for each
query to identify all intersecting database intervals. 
% Formally,
% such a ``searching'' algorithm would return, for each interval $a_i \in A$, the
% set of intervals in $B$ that intersect $a_i$.

However, this apparently straight-forward searching algorithm is complicated by
a subtle, yet vexing detail. If the intervals in $B$ are sorted by their
starting positions, then a binary search of $B$ for the query interval end
position $a_i.end$ will return the interval $b_j \in B$, where $b_j$ is the last
interval in $B$ that starts before interval $a_i$ ends (e.g, interval $e$ in
Fig. 1A).  This would seem to imply that if $b_j$ does not intersect $a_i$, then
no intervals in $B$ intersect $a_i$, and if $b_j$ does intersect $a_i$, then
other intersecting intervals in $B$ could be found by scanning the intervals
starting before $b_j$ in decreasing order, stopping at the first interval that
does not intersect $a_i$.  However, this technique is complicated by the
possibility of intervals that are wholly {\em contained} inside other intervals
(e.g., interval \emph{c} in Fig. 1B). 

An interval $b_j\in B$ is ``contained'' if there exists an interval $b_k \in B$
where $b_k.start \leq b_j.start$ and $b_j.end \leq b_k.end$.  Considering such
intervals, if the interval found in the previous binary search $b_j$ does not
intersect the query interval $a_i$, we cannot conclude that no interval in $B$
intersects $a_i$, because there may exist an interval $b_{j-x} \in B$ where
$b_{j-x}.end > a_i.start$.  Furthermore, if $b_j$ does intersect $a_i$, then the
subsequent scan for other intersecting intervals cannot stop at the first
interval that does not intersect $a_i$; it is possible that some earlier
containing interval intersects $a_i$. Therefore, the scan is forced to continue
until it reaches the beginning of the list. As contained intervals are
typical in genomic datasets, a naive binary search solution is inviable.

\begin{figure}[h]
		\centering
		\includegraphics[width=3in]{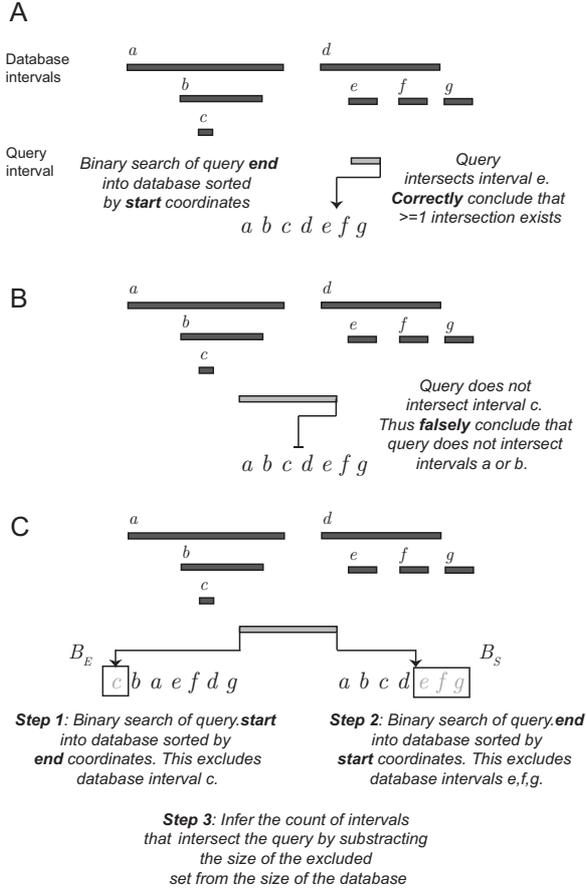}
		\caption{Comparing a naive binary search for interval 
		intersection to the BITS approach. \textbf{A}. Binary
		searches of intervals sorted by start coordinate will
		occasionally identify overlapping intervals. However,  
		contained intervals prevent knowing how far one must scan 
		the database to identify all intersections. \textbf{B}. Contained 
		intervals also cause single binary searches to 
		falsely conclude that no intersections exist for a given query 
		interval.  \textbf{C}. To overcome these limitations, BITS
		uses two binary searches of the database intervals: one
		into a sorted list of end coordinates and the other into a
		sorted list of start coordinates. Each search excludes database
		intervals that \emph{cannot} intersect the query, leaving solely the
		intervals that \emph{must} intersect the query.}
		\label{bitssearching}
\end{figure}

\subsection{Binary Interval Search (BITS) Algorithm}
We now introduce our new Binary Interval Search (BITS) algorithm for solving the
interval set intersection problem.  This algorithm uses binary searches to
identify interval intersections while avoiding the complexities
caused by contained intervals. The key observation underlying BITS 
is that the \emph{size} of the intersection between two sets can be determined
without enumerating each intersection.  For each interval in the query set, two
binary searches are performed to determine the number of intervals in the
database that intersect the query interval.  Each pair of searches is
independent of all others, and thus all searches can be performed in parallel.  

Existing methods define the intersection set based on \emph{inclusion}: that is,
the set of intervals in the interval database $B$ that end after the query
interval $a_i$ begins, and which begin before $a_i$ ends.  However, we have seen 
that contained intervals make it difficult to find this set directly with a 
single binary search.  
% The search for the intersecting intervals can start at the interval in $B$ 
% closest to $a_i$, but must continue to the beginning of $B$ because there is  
% no condition indicating that all intersecting intervals have been found.  
% By restating the intersection
% definition, we are able to determine the size of the intersecting set, which
% provides a terminating condition for the search, so that it may stop once the
% last interval in the intersecting set has been found.

Our algorithm uses a different, but equivalent, definition of interval
intersection based on \emph{exclusion}: that is, by identifying the set of
intervals in $B$ that \emph{cannot} intersect $a_i$, we can infer how many
intervals \emph{must} intersect $a_i$. Formally, we define the set of intervals
$\mathcal{I}(B,a_i) \in B$ that intersect query interval $a_i\in A$ to be the
intervals in $B$ that are neither in the set of intervals ending before (
``left of'', set $\mathcal{L}$ below) $a_i$ begins, nor in the set of intervals
starting after (``right of'', set $\mathcal{R}$ below) $a_i$ ends.  That is:
\vspace{-.75em}
\begin{equation*}
  \begin{split}
      \mathcal{L}(B,a_i) = &\{b\in B| b.end < a_i.start\} \\
      \mathcal{R}(B,a_i) = &\{b\in B| b.start > a_i.end\} \\
      \mathcal{I}(B,a_i) = &B \setminus (\mathcal{L}(B,a_i) \cup \mathcal{R}(B,a_i))
  \end{split}
\end{equation*}
Finding the intervals in $\mathcal{I}(a_i,B)$ for each $a_i\in A$ by taking the
difference of $B$ and the union of $\mathcal{L}(B,a_i)$ and $\mathcal{R}(B,a_i)$
is not efficient.  However, we can quickly find the size of $\mathcal{L}(B,a_i)$
and the size $\mathcal{R}(B,a_i)$, and then \emph{infer} the size of
$\mathcal{I}(B,a_i)$.  With the size of $\mathcal{I}(B,a_i)$, we can directly
answer the decision problem, the counting problem, and the per-interval counting
problems.  The size of $\mathcal{I}(B,a_i)$ also serves as the termination condition for enumerating intersections that was missing in the naive binary search solution.

		%\Comment 
\begin{algorithm}[h]
	\DontPrintSemicolon
	\footnotesize
	\KwIn{Sorted interval starts and ends $B_S$ and $B_E$, query interval $a$}
	\KwOut{Number of intervals $c$ intersecting $a$}
	\BlankLine
	\textbf{Function} \textsc{ICount}$(B_S,B_E,a)$
	\Begin {
		$first \gets \textsc{BSearch}(B_S, a.end)$\;
		$last \gets \textsc{BSearch}(B_E, a.start)$\;
		$c \gets first - last$ 
		\tcc*[f]{$=|B|-(last+(|B|-first))$}

		\Return $c$\;
	}
	\caption{Single interval intersection counter}
	\label{icount}
\end{algorithm}

The BITS algorithm is based upon one fundamental function,
$\textsc{ICount}(B_S,B_E,a_i) = |\mathcal{I}(B,a_i)|$ (Algorithm~\ref{icount}),
which determines the number of intervals in the database $B$ that intersect 
query interval $a_i$.  As shown in Fig. 1C, the database $B$ is split into two 
integer lists $B_S = [b_1.start, b_2.start, \dots, b_M.start]$ and $B_E = 
[b_1.end, b_2.end, \dots, b_M.end]$, which are each sorted numerically in 
ascending order.  Next, two binary searches are performed, 
$last=\textsc{ BSearch}(B_E, a_i.start)$ and $first=\textsc{ BSearch}(B_S,
a_i.end)$.  Since $B_E$ is a sorted list of each interval end coordinate in $B$,
the elements with indices less than or equal to $last$ in $B_E$ correspond to
the set of intervals in $B$ that end \emph{before} $a_i$ starts (i.e., to the
``left'' of $a_i$).  Similarly, the elements with indices greater than or equal
to $first$ in $B_S$ correspond to the set of intervals in $B$ that start
\emph{after} $a_i$ ends (i.e., to the ``right'' of $a_i$).  From these two
values, we can directly infer the size of the intersection set
$\mathcal{I}(B,a_i)$ (i.e., the \emph{count} of intersections in $B$ for $a_i$):
\vspace{-.75em}
\begin{equation*}
	\begin{split}
		|B|-first=&|\mathcal{R}(B,a_i)| \\
		last=&|\mathcal{L}(B,a_i)| \\ 
		|B|-(last+(|B|-first))=&|\mathcal{I}(B,a_i)|
	\end{split}
\end{equation*}

% As mentioned, this problem cannot be solved with a single sorted list because
% contained intervals prevent the total ordering of $B$. If $B$ is sorted by
% interval start coordinates, then the interval end coordinates may be unordered
% and $last$ cannot be found efficiently.  Similarly, if $B$ is sorted by
% interval end, $first$ can not be found efficiently.

Using the subroutine $\textsc{ICount}(B_S,B_E,a_i)$, all four interval set 
intersection problem variants can be solved. Pseudocode for the \emph{decision}, 
\emph{per-interval counting}, and \emph{enumeration} sub-problems can be found 
in the Supplemental Materials.

% \begin{enumerate}

	% \item
	% {\em Decision problem:} Let $c$ be an accumulator variable that is
	% initialized to zero; then for each $a_i \in A$, accumulate $c = c +
	% \textsc{ICount}(B_S,B_E,a_i)$.  If $c\ne0$ return {\em yes}, otherwise
	% return {\em no}.
	{\em The BITS solution to the counting problem.}
	Since BITS operates on arrays of generic intervals 
	($\langle start,end \rangle$), and
	input files are typically chromosomal intervals 
	($\langle chrom,start,end \rangle$), the intervals in each
	dataset are first projected down to a one-dimensional generic interval.  
	This is a straight forward process that adds an offset associated with 
	the size of each chromosome to the start and end of
	each interval.  Once the inputs are projected to interval arrays $A$ and
	$B$, the $\textsc{Counter}$ (Algorithm~\ref{counter}) sets the accumulator
	variable $c$ to zero; then
	for each $a_i \in A$, accumulates $c = c + \textsc{ICount}(B_S,B_E,a_i)$.
	The total count $c$ is returned.
	\begin{algorithm}[h]
		\DontPrintSemicolon
		\footnotesize
		\KwIn{Database interval array $B$ and query interval array $A$}
		\KwOut{Number of intersections $c$ between $A$ and $B$}
		\BlankLine
		\textbf{Function} \textsc{Counter}$(A,B)$
		\Begin {
			$B_S \gets [b_1.start, \dots, b_{|B|}.start]$\;
			$B_E \gets [b_1.end, \dots, b_{|B|}.end]$\;
			\textsc{Sort}($B_S$)\;
			\textsc{Sort}($B_E$)\;
			$c \gets 0$\;
			\For{$i \gets 1$ \KwTo $|A|$} {
				$c \gets c + \textsc{ICount}(B_S,B_E,A[i])$
			}
			\Return $c$\;
		}
		\caption{Interval intersection counter}
		\label{counter}
	\end{algorithm}

\subsection{Time Complexity Analysis}

The time complexity of BITS is $O( (|A|+|B|) \log |B|)$, which can
be shown to be optimal by a straight-forward reduction to element uniqueness
(known to be $\Theta(N\log N)$~\citep{misra1982}).  To compute
$\textsc{ICount}(B_S,B_E,a_i)$ for each $a_i$ in $A$, the interval set $B$ is
first split into two sorted integer lists $B_S$ and $B_E$, which requires $O(|B|
\log |B|)$ time.  Next, each instance of $\textsc{ICount}(B_S,B_E,a_i)$ searches
both $B_S$ and $B_E$, which consumes $O(|A| \log |B|)$ time.  For the counting
problems, combining the results of all $\textsc{ICount}(B_S,B_E,a_i)$ instances
into a final result can be accomplished in $O(|A|)$ time.  

% The total complexity of the counting problems is therefore $O((|A| + |B|) \log
% |B|)$.  The enumeration problem requires additional steps to scan the
% intervals in $B_S$.  In the best case scenario, each scan requires
% $\textsc{ICount}(B_S,B_E,a_i)$ extra steps, to a total of $O(|B| \log |B| +
% C)$ time, where is $C$ is the number of intersections.  However, contained
% intervals can cause the scan to process more than $C$ elements.  If there
% exists some $a_i$ that intersects $b_{j}$ and $b_{j-2}$, but not $b_{j-1}$
% (i.e., $b_{j-2}$ contains $b_{j-1}$), then the enumeration scan will consider
% one extra element, namely $b_{j-1}$.  In the pathological case, $b_0$ contains
% intervals $\{b_1, \dots, b_N\} \in B$, $a_i$ intersects $b_0$, and $a_i$
% starts after interval $b_N$ ends.  This scenario would cause the enumeration
% scan to consider all the elements in $B$.  If all $a_i$ in $A$ are
% pathological, then each scan would required $|B|$ extra steps, to a total of
% $O(|B| \log |B| + |A||B|)$ time.

%%%%%%%%%%%%%%%%%%%%%%%%%%%%%%%%%%%%%%%%%%%%%%%%%%%%%%%%
% METHODS: Parallel BITS
%%%%%%%%%%%%%%%%%%%%%%%%%%%%%%%%%%%%%%%%%%%%%%%%%%%%%%%%
\subsection{Parallel BITS}

Performing a single operation independently on many different inputs is a
classic parallelization scenario.  When based on the subroutine
$\textsc{ICount}(B_S,B_E,a)$, which is independent of all
$\textsc{ICount}(B_S,B_E,x)$ for intervals $x$ in the query set where $a \neq
x$, counting interval intersections is a {\em pleasingly parallelizable} problem
that easily maps to a number of parallel architectures.

%No need to spell out CUDA. Explain that in SIMD architectures, threads operate
%in lock-step (right now, you mention SIMD and lock-step but not together).
%Also clarify that the GPU has multiple SIMD processing units, so only the batch
%of threads on one SIMD unit need to operate in lock-step.

NVIDIA's CUDA is a single instruction multiple data (SIMD) architecture that
provides a general interface to a large number of parallel GPUs.  The GPU is
organized into multiple SIMD processing units, and the processors within a unit
operate in lock-step.  The BITS
algorithm is especially well suited for the this 
architecture for a number of reasons.  First, CUDA is optimized to handle large 
numbers of threads. By assigning each thread one instance of 
$\textsc{ICount}(B_S,B_E,a)$, the number of threads will be proportional to the 
input size.  CUDA threads also execute in lock-step and any divergence between 
threads will cause reduced thread utilization.  While there is some divergence
in the depth of each binary search performed by $\textsc{ICount}(B_S,B_E,a)$, it
has an upper bound of $O(log |B|)$.  Outside of this divergence
$\textsc{ICount}(B_S,B_E,a)$ is a classic SIMD operation~\citep{kirk2010}.
Finally, the only data structure required for this algorithm is a sorted array, 
and thanks to years of research in this area, current GPU sorting algorithms can 
sort billions of integers within seconds~\citep{merrill2011,satish2009}.

%%%%%%%%%%%%%%%%%%%%%%%%%%%%%%%%%%%%%%%%%%%%%%%%%%%%%%%%
% RESULTS
%%%%%%%%%%%%%%%%%%%%%%%%%%%%%%%%%%%%%%%%%%%%%%%%%%%%%%%%
\section{Results}

%%%%%%%%%%%%%%%%%%%%%%%%%%%%%%%%%%%%%%%%%%%%%%%%%%%%%%%%
% RESULTS: Comparing sequential BITS to existing approaches
%%%%%%%%%%%%%%%%%%%%%%%%%%%%%%%%%%%%%%%%%%%%%%%%%%%%%%%%
\subsection{Comparing BITS to extant sequential approaches}
We implemented a sequential version of the BITS algorithm ("BITS-SEQ") as a 
stand-alone C++ utility. Here we assess the performance of this implementation 
relative to BEDTOOLS \texttt{intersect} and UCSC Genome Browser's (``UCSC'')
\citep{kent2002} \texttt{bedIntersect} utilities (see Supplemental Materials for
details).  We compare the performance of each tool for \emph{counting} the total
number of observed intersections between sets of intervals of varying sizes
(Figure 2). The comparisons presented are based on sequence alignments for the 
CEU individual NA12878 by the 1000 Genomes Project~\citep{durbin2010}, as well 
as RefSeq exons. Owing
to the different data structures used by each algorithm, the relative
performance of each approach may depend on the genomic distribution of intervals
within the sets. As discussed previously, tree-based solutions
that place intervals into hierarchical bins may perform poorly when intervals
are unevenly distributed among the bins. We tested the impact of differing
interval distributions on algorithm performance by randomly sampling 1 and 10
million alignment intervals from both whole-genome and exome-capture datasets
for NA12878 (see Supplemental Materials). Each algorithm was evaluated
considering three different interval intersection scenarios. First, we tested 
intervals from \emph{different} distributions by comparing the intersection 
between exome-capture alignments and whole-genome alignments. Since each set
has a large number of intervals and a different genomic 
distribution, we expect a small (relative to the total set size) number of 
intersections. We also tested a \emph{uniform} distribution by counting 
intersections between Refseq exons and whole-genome sequencing alignments.
Here each interval set is, for the most part, evenly distributed throughout
the genome; thus, we expect each exon to intersect roughly the same number
of sequencing intervals, and a large number of sequencing intervals will not
intersect an exon. Lastly, we assessed a \emph{biased} intersection
distribution between exons and exome-capture alignments. By design, 
exome sequencing experiments intentionally focus collected
DNA sequences to the coding exons. Thus, the vast majority of sequence intervals 
will align in exonic regions. In contrast to the previous scenario, nearly every 
exon interval will have a large number of sequence interval intersections, and 
nearly all sequencing intervals will intersect an exon.

\begin{figure*}[btp]
	\centering
	\includegraphics[width=6.5in]{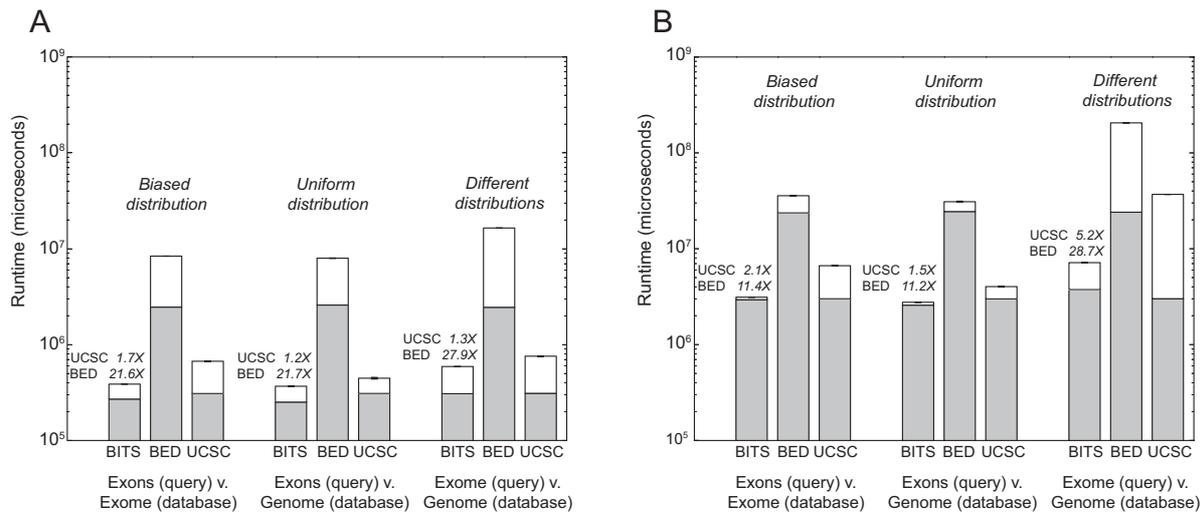}
	\caption[]{Run times for counting intersections with 
	BITS, BEDTools, and UCSC ``Kent source''.  \textbf{A}. Run times for
databases of 1 million alignment intervals from each interval distribution.
	\textbf{B}.  Run times for databases of 10 million alignment intervals from
	each interval distribution.  Bars reflect the mean run
	time from five independent experiments and error bars describe the standard
	deviation.  Gray bars reflect the run time consumed by
	data structure construction, while white bars
	are the time spent counting intersections.  Above each BITS
	execution time we note the speed increase relative to BEDTools and 
	``Kent source'', respectively. ``Exons'' represents 400,351 RefSeq exons
	(autosomal and X, Y) for the human genome (Build 37).  BED = BEDTools; UCSC
	= ``Kent source''.}
	\label{seq-counting}
\end{figure*}

% \begin{enumerate}
%   \item {\em Intervals from different distributions}:
%   the intersection between exome-capture alignments and whole-genome
%   alignments. Since the interval sets have a large number of intervals and
%   each set has a different genomic distribution, we expect a small number of
%   intersections.
%   
%   \item {\em Uniform interval distribution}:
%   the intersection between Refseq exons and genome-wide sequencing data.
%   Since each interval set is, for the most part, evenly distributed throughout
%   the genome, we expect that each exon will intersect roughly the same number
%   of sequencing intervals, and a large number of sequencing intervals will not
%   intersect an exon.
% 
%   \item {\em Biased interval distribution}:
%   the intersection between exons and exome-capture alignments. By design, 
%   exome sequencing experiments intentionally bias DNA sequences to the coding
%   exons. Thus, the vast majority of sequence intervals will align in exonic
%   regions. In contrast to the previous scenario, nearly every exon interval
%   will have a large number of sequence interval intersections, and nearly all
%   sequencing intervals will intersect an exon.
% \end{enumerate}
        
\subsubsection{BITS excels at counting intersections.}

In all three interval distribution scenarios, the sequential version of BITS had
superior runtime performance for counting intersections.  BITS was between 11.2
and 27.9 times faster than BEDTools and between 1.2 and 5.2 times faster than
UCSC (Figure~\ref{seq-counting}). This behavior is expected; whereas the
BEDTools and UCSC tree-based algorithms must \emph{enumerate} intersections
to derive the \emph{count}, BITS \emph{infers} the intersection count by 
exclusion without enumeration.

\subsubsection{BITS excels at large intersections and biased distributions.}
The relative performance gains of the BITS approach are enhanced for very large
datasets (Figure 2B).  Since tree-based methods have a fixed number of bins, and 
searches require a linear scan of each associated bin,
the number of intervals searched grows linearly with respect to the input size.
% (albeit with a very small coefficient).  
In the worst-case where all intervals
are in a single bin, a search would scan the entire input set.  In contrast, 
BITS employs binary searches so the number of operations is proportional 
to $\log$ of the input size, regardless of the input distribution.

%The relative performance gains of the BITS approach are enhanced for very large
%datasets (Figure 2B), as the tree-based methods will, on average, have a large
%number of intervals in each queried bin. In contrast, since BITS is based on
%binary search ($O(\log |B|)$ complexity), its performance scales more linearly
%with respect to dataset size.

Similarly, exome-capture experiments yield biased distributions of intervals 
among the UCSC bins. Consequently, most bins in tree-based methods will contain 
no intervals, while a small fraction contain
many intervals. When the query intervals have the same bias, the overhead of the
UCSC algorithm is more onerous, as a small number of bins are queried and each
queried bin contains many intersecting intervals that must be enumerated in
order to count overlaps. As the BITS algorithm is agnostic to the interval
distributions, it will outperform the UCSC algorithm (Figure 2A, 2B) for common
genomic analyses such as ChIP-seq and RNA-seq, especially given the massive 
size of these datasets.

\subsection{Applications for Monte Carlo Simulations}

Identifying statistically significant relationships between sets of genome
intervals is fundamental to genomic research. However, owing to our complex 
evolutionary history, different classes of genomic features have distinct
genomic distributions, and as such, testing for significance can be challenging. 
One widely-used, yet computationally intensive alternative solution is the 
use of Monte Carlo simulations that compare \emph{observed} interval 
relationships to an \emph{expectation} based on randomization. 
All aspects of the BITS algorithm are particularly well suited for Monte Carlo
(MC) simulations measuring relationships between interval sets. As described,
all intersection algorithms begin detecting intersections between two interval
sets by setting up their underlying data structures (e.g., trees or arrays). The 
BITS setup process
involves mapping each interval from the two-dimensional chromosomal interval
space (i.e., chromosome and start/end coordinates) to a one dimensional integer
interval space (i.e., start/end coordinates ranging from 1 to the total genome
size). Once the intervals are mapped, arrays are sorted by either start or
end coordinates.  In contrast, the UCSC setup places each interval into a hash 
table.  As shown in Figure~\ref{seq-counting}, data
structure setup is a significant portion of the total runtime for all
approaches. 

However, in the case of many MC simulation rounds, where a uniformly distributed
random interval set is generated and placed into the associated data structure,
the setup step is faster in BITS, whereas the setup time remains constant in
each simulation round for UCSC.  For BITS, the mapping step is skipped in all
but the the first round and in each simulation round only an array of random
starts must be generated. The result is a 6x speedup for MC rounds over the
cost of the initial intersection setup.  For UCSC, both the chromosome and the
interval start position must be generated and then placed into the hash table
with no change in execution time.

This speedup in BITS is extended on parallel platforms, where the independence
of each intersection is combined with efficient parallel random number
generation algorithms ~\citep{tzeng2008} and parallel sorting
algorithms~\citep{merrill2011,satish2009}.  Monte Carlo simulations have obvious
task parallelism since each round is independent.  BITS running on CUDA 
("BITS-CUDA") goes a step further and exposes fine-grain parallelism in both the 
setup step, with
parallel random number generation and parallel sorting, and the intersection
step where hundreds of intersections execute in parallel.  The improvement is
modest for a single intersection (only parallel sorting can be applied to the
setup step) where BITS-CUDA is 4x faster than sequential BITS and 40x faster
than sequential UCSC.  However, as the number of MC rounds grows performance
improves dramatically.  At 10,000 MC rounds and 1e7 intervals, BITS-CUDA is 267x
faster than sequential BITS and 3,414x faster than sequential UCSC.  An
improvement of this scale allows MC analyses for thousands of experiments 
(e.g., 25,281 pairwise comparisons in Section~\ref{hm:section}).

We demonstrate the improved performance of BITS over UCSC for Monte Carlo
simulations for measuring the significance of the overlaps between interval sets
in Table~\ref{table:avge}.  As both the number of MC rounds and the size of
the dataset grows, the speedup of both sequential BITS and BITS-CUDA increases
over UCSC.  For the largest comparison (1e7 intervals and 10,000 iterations),
BITS-SEQ is 12x faster that UCSD, and BITS-CUDA is 267x
faster than BITS-SEQ and 3,414x faster than sequential UCSC.

% NOTE: moved to figure legend
% Due to long runtimes, timings for sequential BITS (BITS-SEQ) and UCSC with
% set sizes 1e6 and 1e7 and iterations 1,000 and 10,000 (in italics) were
% estimated from their 1e5 times assuming a linear scaling.

\begin{center}
	\begin{table}[h!b!p!]
	\caption{Runtime (seconds) comparison for Monte Carlo simulations. 
	Timings in italics were extrapolated owing to long run times.}
	\begin{tabular}{l l c c c c}
	\multicolumn{2}{c}{} & \multicolumn{4}{c}{Number of MC iterations} \\
	\cline{3-6}
	Size & Tool & 1 & 100 & 1000 & 100000 \\
	\hline
	1e5 & BITS-CUDA & 0.73 & 1  & 4   & 28 \\
		& BITS-SEQ  & 0.41 & 7  & 68  & 680 \\
		& UCSC      & 0.17 & 14 & 138 & 1,381 \\
	1e6 & BITS-CUDA & 2 & 3    & 1       & 103 \\
		& BITS-SEQ  & 5 & 120  & \emph{1,200} & \emph{12,000} \\
		& UCSC      & 6 & 878  & \emph{8,780} & \emph{87,800} \\
	1e7 & BITS-CUDA & 14  & 22    & 97            & 835 \\
		& BITS-SEQ  & 66  & 2,235  & \emph{22,350}  & \emph{223,500} \\
		& UCSC      & 568 & 28,508 & \emph{285,080} & \emph{2,850,800} \\
	
	\hline
	\end{tabular}
	\label{table:avge}
	\end{table}
\end{center}

\vspace{-2em}
\subsection{Uncovering novel genomic relationships.}
\label{hm:section}
The efficiency of BITS for Monte Carlo applications on GPU architectures
provides a scalable platform for identifying novel relationships between
large scale genomic datasets. To illustrate BITS-CUDA's potential
for large-scale data mining experiments, 
we conducted a screen for significant genomic co-localization among
159 genome annotation tracks using Monte Carlo simulation (see
Supplemental Materials). This analysis was based upon functional annotations 
from the ENCODE project~\citep{encode2007} for the GM12878, H1-hESC, and K562 
cell lines, including assays for 24 transcriptions factors 
(often with replicates), 8 histone
modifications, open chromatin, and DNA methylation. We also included diverse 
genome annotations from the UCSC genome browser (e.g, repeats, 
genes, and conserved regions).

Using BITS-CUDA, we measured the log2 ratio of the observed and expected number
of intersections for each of the 25,281 (i.e., 159*159) pairwise 
dataset relationships using 1e4 Monte Carlo simulations (Figure 3).
As expected, this analysis revealed that 1) the genomic locations 
for the same functional element are largely consistent across 
replicates and cell types, 2) methylated and semi-methylated regions
are similar across cell types, and 3) most functional assays were
anti-correlated with genomic repeats (e.g., microsatellites)
owing to sequence alignment strategies that exclude repetitive genomic
regions. Perhaps not surprisingly, this unbiased screen also revealed
intriguing patterns.  First, the strong enrichment among all
transcription factors (TF) assays suggests that a subset of TF
binding sites are shared among all factors. This observation is 
consistent with previous descriptions of ``hot regions''
~\citep{gerstein2010}. In addition, there is a significant, 
specific, and unexplained enrichment among the Six5 transcription factor
and segmental duplications. 

Pursuing the biology of these relationships is beyond the 
scope of the current manuscript; however, we emphasize that the ability 
to efficiently conduct such large-scale screens facilitates novel 
insights into genome biology. This analysis presented a tremendous
computational burden made feasible by the facility with which
the BITS algorithm could be applied to GPU architectures. Indeed, each
iteration of our Monte Carlo simulation tested for
intersections among 4 billion intervals among the 25 thousand datasets,
yielding over 44 trillion comparisons for the entire simulation. Whereas
this simulation took just over 6 days (9,069 minutes) on a single
computer with one GPU card, we estimate that it would take at least 
112 traditional processors to conduct the same analysis using 
traditional approaches such as the UCSC tools or BEDTools.

\begin{figure*}[btp]
        \includegraphics[width=7in,height=7in]{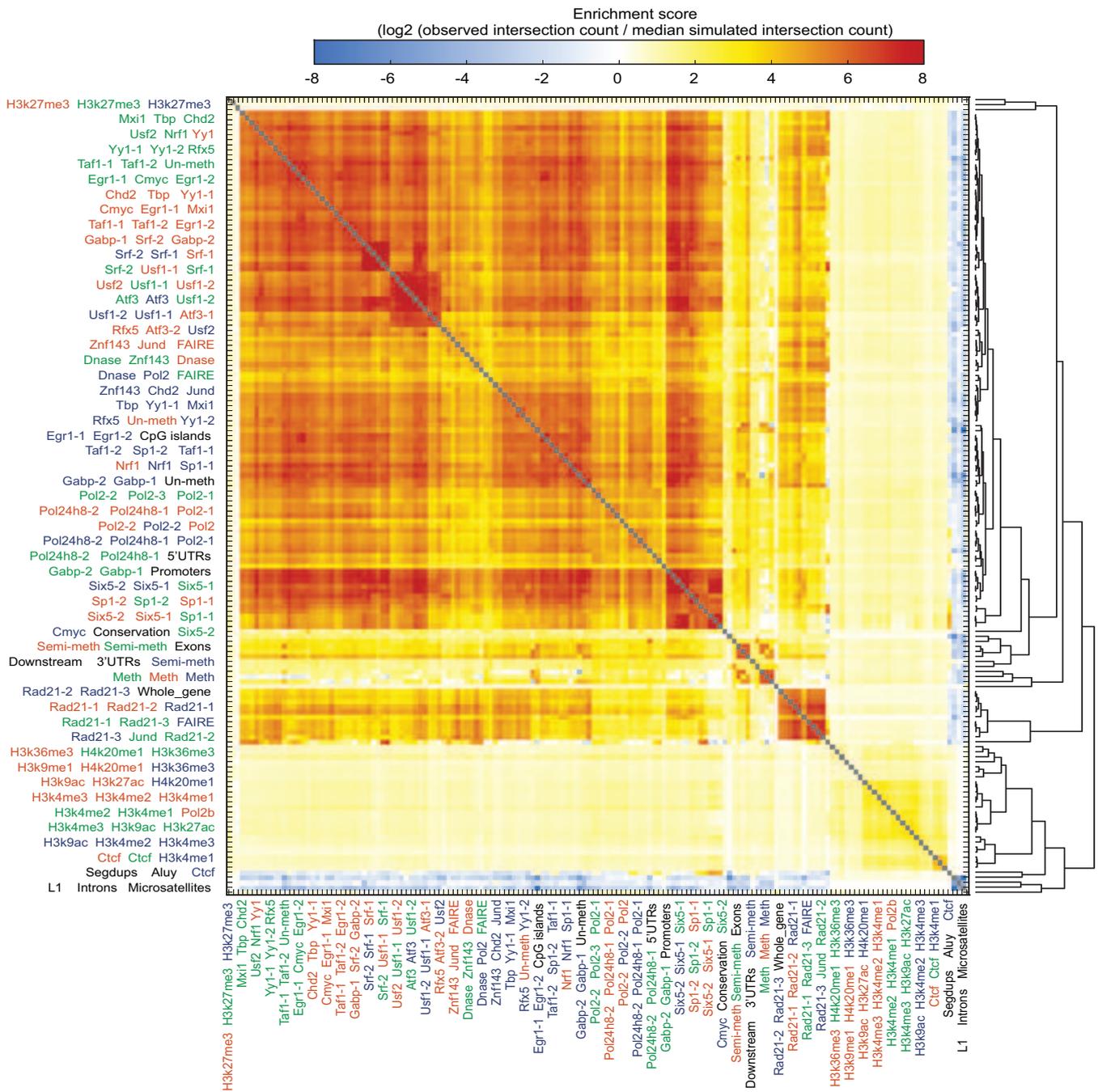}
        \caption[]{BITS-CUDA measurements of spatial correlations
among 159 genome features from the ENCODE project and from the UCSC
Genome Browser. For each comparison, we show an enrichment score
reflecting the log2 ratio of the observed count of intersections over
the median count of intersections from 10000 MC simulations. Each set of three
labels on the X and Y axes correspond to three consecutive rows or columns,
respectively. Assays from the GM12878 cell line are in green, H1-hESC in blue,
and K562 in red.  Annotation tracks from the UCSC Genome Browser are in black.}
\end{figure*}

%%%%%%%%%%%%%%%%%%%%%%%%%%%%%%%%%%%%%%%%%%%%%%%%%%%%%%%%
% CONCLUSION
%%%%%%%%%%%%%%%%%%%%%%%%%%%%%%%%%%%%%%%%%%%%%%%%%%%%%%%%

\vspace{-2em}
\section{Conclusion}
We have developed a novel algorithm for interval intersection that
is uniquely suited to scalable computing architectures such as GPUs.
Our algorithm takes a new approach to counting intersections: 
unlike existing methods that must enumerate intersection in order to 
derive a count, BITS uses two binary searches to directly infer the 
count by excluding intervals that \emph{cannot} intersect one another. 

We have demonstrated that a sequential implementation of BITS outperforms 
existing tools and illustrate that, because it is based on binary searches
(which have predictable complexity), BITS is task efficient and is thus highly 
parallelizable. As such, we show that a GPU implementation of BITS (BITS-CUDA)
is a superior solution for Monte Carlo analyses of statistical relationships 
between sets of genome intervals, since observed intersections among many sets 
must be compared to thousands of randomized simulations.
% Using a GPU implementation of BITS,
% we highlighted the data mining potential of our approach by 
% exploring relationships among 161 genome annotations and assays of 
% functional elements from the ENCODE project.

Given the efficiency with which the BITS algorithm counts intersections,
it is also perfectly suited to many fundamental genomic analyses
including RNA-seq transcript quantification, ChIP-seq peak detection, and 
searches for copy-number and structural variation. Moreover, the 
functional and regulatory data produced by projects such as ENCODE
have driven the development of new approaches~\citep{favorov2012} 
to measuring relationships among genomic features in order to reveal yet 
undetected insights into genome biology. We recognize the importance of 
scalable approaches to detecting such relationships and anticipate that 
our new algorithm will foster new genome mining tools for the 
genomics community.
\vspace{-2em}
\section*{ACKNOWLEDGEMENTS}
We are grateful to Anindya Dutta for helpful discussions throughout the
preparation of the manuscript and to Ryan Dale for providing scripts that
aided in the analysis and interpretation of our results.

\paragraph{Funding\textcolon} This research was supported by an NHGRI award to AQ (NIH 1R01HG006693-01).
\vspace{-2em}
        \bibliographystyle{natbib}
        \bibliography{bioinformatics}
\end{document}